\documentclass[conference]{IEEEtran}
\usepackage{xcolor}
\usepackage{balance}
\usepackage{cite}
\usepackage{multirow}
\usepackage[pdftex]{graphicx}
\usepackage{amsmath,amssymb,amsfonts}
\usepackage{textcomp}
\usepackage[caption=false,font=footnotesize]{subfig}
\usepackage[nolist]{acronym}
\usepackage[capitalize]{cleveref}
\usepackage{nicefrac}
\usepackage{bm}
\usepackage{graphicx}  
\crefname{table}{Tab.}{Tabs.}
\Crefname{table}{Tab.}{Tabs.}
\crefformat{equation}{(#2#1#3)}

\newcommand\norm[1]{\left\lVert#1\right\rVert}

\newcommand{\Estimate}[2]{\widetilde{\mathbf{#1}}^{\left( {\rm #2} \right)} }
\newcommand{\EstimateBeam}[2]{\widetilde{\mathbf{#1}}_{\rm B}^{\left( {\rm #2} \right)} }
\newcommand{\HatBeam}[2]{\widehat{\mathbf{#1}}_{\rm B}^{\left( {\rm #2} \right)} }
\newcommand{\BarX}[2]{\overline{\mathbf{#1}}^{\left( {\rm #2} \right)} }
\newcommand{\BarBeamX}[2]{\overline{\mathbf{#1}}_{\rm B}^{\left( {\rm #2} \right)} }
\newcommand{\NoLine}[2]{\mathbf{#1}^{\left( {\rm #2} \right)} }
\newcommand{\NoLineBM}[2]{\bm{#1}^{\left( {\rm #2} \right)} }
\newcommand{\TRx}[3]{#1_{\rm #2}^{\left( {\rm #3} \right)} }

\newcommand{\vectorize}[1]{ {\rm vec} \left( #1 \right) }
\newcommand{\unvectorize}[1]{ {\rm unvec} \left( #1 \right) }
\newcommand{\abs}[1]{ \left| #1 \right| }

\newcommand{\NormalBrackets}[1]{ \left( #1 \right) }
\newcommand{\SquareBrackets}[1]{ \left[ #1 \right] }
\newcommand{\CurlyBrackets}[1]{ \left\{ #1 \right\} }

\begin{document}
\begin{acronym}

\acro{5G}[5G]{fifth generation}
\acro{3GPP}[3GPP]{3rd Generation Partnership Project}

\acro{ADC}[ADC]{analog-to-digital converter}
\acro{AoD}[AoD]{angle-of-departure}
\acro{AoA}[AoA]{angle-of-arrival}
\acro{AWG}[AWG]{Arbitrary Waveform Generator}
\acro{AWGN}[AWGN]{additive white Gaussian noise}

\acro{BER}[BER]{bit error ratio}

\acro{CDF}[CDF]{cumulative distribution function}
\acro{CIR}{channel impulse response}
\acro{CW}[CW]{Continuous Wave}
\acro{CSIT}{channel state information at the transmitter}
\acro{CSI}{channel state information}
\acro{CTF}{channel transfer function}
\acro{CTF}[CTF]{channel transfer function}

\acro{DAC}[DAC]{digital-to-analog converter}
\acro{DFT}[DFT]{discrete Fourier transform}
\acro{DPSS}[DPSS]{discrete prolate spheroidal sequences}
\acro{DR}[DR]{dynamic range}
\acro{DSD}{Doppler power spectral density}

\acro{EVD}[EVD]{eigenvalue decomposition}
\acro{eMBB}[eMBB]{enhanced mobile broadband}

\acro{FDD}{frequency-division duplex}
\acro{FPGA}[FPGA]{Field Programmable Gate Array}
\acro{FSPL}[FSPL]{free space path loss}

\acro{GRU}{Gated Recurrent Unit}

\acro{HPBW}[HPBW]{half-power beamwidth}
\acro{HST}[HST]{High-Speed Train}

\acro{IC}[IC]{Integrated Circuit}
\acro{ICI}[ICI]{Inter-Carrier Interference}
\acro{IDFT}[IDFT]{inverse discrete Fourier transform}
\acro{IF}[IF]{Intermediate Frequency}
\acro{IFFT}{Inverse Fast Fourier Transform}
\acro{ISI}[ISI]{Inter-Symbol Interference}
\acro{ITS}[ITS]{intelligent transportation systems}
\acro{ISI}[ISI]{Inter-Symbol Interference}

\acro{LMMSE}[LMMSE]{linear minimum mean square error}
\acro{LO}[LO]{Local Oscillator}
\acro{LOS}[LOS]{line-of-sight}
\acro{LSF}[LSF]{local scattering function}
\acro{LSTM}{long short-term memory}
\acro{LS}[LS]{least-squares}

\acro{MAC}[MAC]{medium access control}
\acro{MIMO}[MIMO]{multiple-input multiple-output}
\acro{ML}{machine learning}
\acro{MMW}[mmWave]{millimeter wave}
\acro{MRC}[MRC]{maximal ratio combining}
\acro{MSE}[MSE]{mean squared error}
\acro{MUSIC}[MUSIC]{multiple signal classification}

\acro{NR}[NR]{new radio}
\acro{NLOS}[NLOS]{non-line-of-sight}

\acro{OFDM}[OFDM]{orthogonal frequency-division multiplexing}
\acro{OOBA-BD}[OOBA-BD]{out-of-band aided beam-domain}
\acro{OBABE}[OBABE]{out-of-band aided beam-domain estimation}

\acro{PDP}{power delay profile}
\acro{PCB}[PCB]{Printed Circuit Board}

\acro{RF}[RF]{radio frequency}
\acro{RMS}[RMS]{root-mean-square}
\acro{RMSE}[RMSE]{root mean squared error}
\acro{RNN}{recurrent neural networks}

\acro{QAM}[QAM]{quadrature amplitude modulation}

\acro{SE}[SE]{spectral efficiency}
\acro{SISO}[SISO]{single-input single-output}
\acro{SIMO}[SIMO]{single-input multiple-output}
\acro{SINR}[SINR]{signal-to-interference-and-noise ratio}
\acro{SMD}[SMD]{Surface Mount Device}
\acro{SNR}[SNR]{signal-to-noise ratio}
\acro{SVD}[SVD]{singular value decomposition}
\acro{SVM}[SVM]{support vector machine}

\acro{TDD}{time-division duplex}
\acro{TDL-A}[TDL-A]{tapped delay line A}
\acro{TDL-D}[TDL-D]{tapped delay line D}

\acro{URLLC}[URLLC]{ultra-reliable low-latency communication}
\acro{ULA}[ULA]{uniform linear array}

\acro{V2X}[V2X]{vehicle-to-everything}
\acro{VNA}[VNA]{vector network analyzer}

\end{acronym}

\title{Beam-Domain Channel Estimation for mmWave MIMO using Sub-6\,GHz Out-of-Band Information}

\author{\IEEEauthorblockN{
Faruk Pasic,
Mariam Mussbah,
Stefan Schwarz and
Markus Rupp
}%

\IEEEauthorblockA{
Institute of Telecommunications, TU Wien, Vienna, Austria}
\IEEEauthorblockA{faruk.pasic@tuwien.ac.at}
}

\IEEEoverridecommandlockouts 

\makeatletter
\def\thanks#1{\protected@xdef\@thanks{\@thanks
        \protect\footnotetext{#1}}}
\makeatother

\maketitle

\begin{abstract}
Future wireless \ac{MIMO} systems will integrate sub-6\,GHz and \ac{MMW} bands to support high data rates for latency-critical applications.
The coexistence of these bands enables the use of reliable out-of-band sub-6\,GHz information to assist \ac{MMW} link establishment, particularly for channel estimation.
In this paper, we propose a novel beam-domain channel estimation method for \ac{MMW} \ac{MIMO} that leverages sub-6\,GHz beam-domain information to improve estimation accuracy.
We compare the proposed method with conventional in-band methods. 
Simulation results show that our proposed method outperform existing baselines in terms of spectral efficiency, in both \ac{LOS} and \ac{NLOS} scenarios.
\end{abstract}
\vskip0.5\baselineskip

\acresetall

\section{Introduction} \label{sec:introduction}
Mainstream wireless systems mainly operate within the sub-6\,GHz frequency range.
However, these bands are increasingly unable to satisfy the growing demand for higher data rates, due to limited available spectrum.
In contrast, \ac{MMW} bands (24\,GHz -- 300\,GHz) provide substantially larger bandwidth, which enable significantly higher data rates~\cite{Rappaport2013}.
Therefore, \ac{MMW} technology is widely considered a key enabler for next-generation communications~\cite{Molisch2025, Hammoud2025_TVT}.

Despite their advantages, \ac{MMW} systems also introduce several challenges.
Owing to their much shorter wavelengths compared to sub-6\,GHz signals, \ac{MMW} transmissions experience significantly higher propagation losses~\cite{Rappaport2013}.
To mitigate this frequency-dependent path loss and maintain a reliable link margin, \ac{MMW} systems typically employ large antenna arrays~\cite{Heath2016}. 
This, in turn, enables the use of \ac{MIMO} techniques with highly directional beamforming and spatial multiplexing.
However, exploiting large antenna arrays introduces the additional challenge of link establishment.
This process, which relies on beam selection or channel estimation, becomes particularly challenging at \ac{MMW} frequencies due to the inherently low pre-beamforming \ac{SNR}~\cite{Heath2016}.

In addition, \ac{MMW} systems are being deployed in conjunction with sub-6\,GHz systems to support multi-band operation and enhance coverage~\cite{Hofer2025}.
Due to their lower propagation losses, sub-6\,GHz signals achieve higher pre-beamforming \ac{SNR}.
Furthermore, recent multi-band measurement studies have revealed that both frequency ranges exhibit comparable multipath behavior, particularly in the delay and angular domains~\cite{Seun2018, Hofer2024, Miao2023, Pasic2025_OJCOMS}.
Hence, these cross-band similarities can be exploited to improve \ac{MMW} link establishment by utilizing reliable out-of-band information from the sub-6\,GHz band.

So far, variety methods have been proposed to exploit sub-6\,GHz out-of-band information to support \ac{MMW} \ac{MIMO} link establishment~\cite{Nitsche2015, Ali2019, Sim2020}.
However, these methods mainly target analog or hybrid beamforming architectures, which rely on sequential beam sweeping and incur significant time overhead, in particular when considering large antenna arrays on both link ends.
In contrast, fully digital architectures enable parallel beamforming, substantially reducing time overhead and making them attractive for latency-critical applications~\cite{Liu2020}.
Despite these advantages, the use of out-of-band information in fully digital \ac{MMW} systems remains underexplored.
Existing work~\cite{Pasic2025_TCOM, Pasic2025_INFOCOM} shows performance gains primarily under \ac{LOS} conditions, with limited improvements in \ac{NLOS} scenarios.
Therefore, advanced solutions are needed to ensure reliable performance in both propagation environments.

\textbf{Contribution:}
In this paper, we propose a novel channel estimation method for \ac{MMW} \ac{MIMO} systems.
The proposed method operates in the beam domain and leverages beam-domain estimates from the sub-6\,GHz band to improve estimation accuracy in the \ac{MMW} band.
We evaluate its performance via Monte-Carlo simulations in terms of \ac{SE}, considering both \ac{LOS} and \ac{NLOS} propagation conditions characterized by the Rician $K$-factor.

\textbf{Organization:}
The remainder of this paper is organized as follows.
In~\cref{sec:system_model}, we present the system model, while in~\cref{sec:methods}, we introduce the considered channel estimation methods.
\cref{sec:simulations} provides a performance evaluation based on simulations and \cref{sec:conclusion} concludes the paper.

\textbf{Notation:}
The superscript $\left( \cdot \right) ^{\left( \rm b \right)}$ indicates frequency-band-dependent quantities, where ${\rm b} \in \{ {\rm s}, {\rm m} \}$.
Here, ${\rm s}$ refers to the sub-6\,GHz band and ${\rm m}$ to the \ac{MMW} band.
Scalars are denoted by lowercase letters $x$, column vectors by bold lowercase letters ${\mathbf x}$ and matrices by bold uppercase letters ${\mathbf X}$.
The $i$-th row and $j$-th column of ${\mathbf X}$ are represented by $\mathbf{X}_{i, :}$ and $\mathbf{X}_{:, j}$, respectively.
The Kronecker product is denoted by $\otimes$ and matrix vectorization by $\mathrm{vec}\left( \cdot \right)$.
The superscripts, $\left( \cdot \right) ^{\rm T}$,  $\left( \cdot \right) ^*$ and $\left( \cdot \right) ^{\rm H}$ correspond to transpose, complex conjugation and Hermitian transpose, respectively.
The Frobenius norm is written as $\lVert \cdot \rVert_F$, while $\lVert \cdot \rVert$ denotes the Euclidean norm.
Finally, $\delta [\cdot]$ represents the discrete delta function.

\section{System Model} \label{sec:system_model}
\begin{figure}[t]
    \centering
    {\includegraphics[width=\columnwidth]{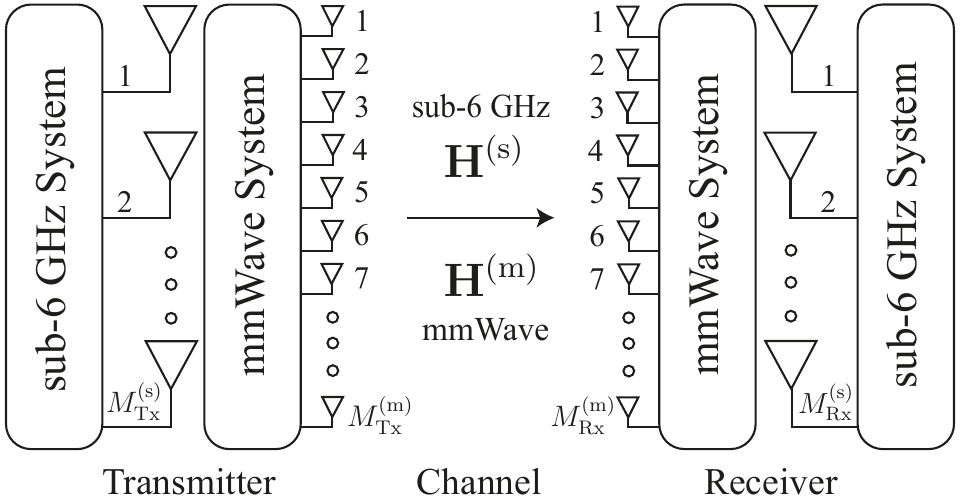}}
    \caption{
    The considered multi-band MIMO system employs co-located antenna arrays that operate in both sub-6\,GHz and mmWave frequency bands.
    }
    \label{fig:system_model}
\end{figure}

We consider a point-to-point multi-band \ac{MIMO} system where sub-6\,GHz and \ac{MMW} subsystems operate simultaneously under far-field conditions. 
The transmitter is equipped with $\TRx{M}{Tx}{s}$ sub-6\,GHz and $\TRx{M}{Tx}{m}$ \ac{MMW} antenna elements. 
Similarly, the receiver employs $\TRx{M}{Rx}{s}$ sub-6\,GHz and $\TRx{M}{Rx}{m}$ \ac{MMW} antennas, as illustrated in~\cref{fig:system_model}.
The sub-6\,GHz and \ac{MMW} arrays are co-located, aligned \acp{ULA}\footnote{The analysis in~\cite{Pasic2026_iWAT} shows that co-located and aligned sub-6\,GHz and \ac{MMW} \acp{ULA} are practically feasible, with only minimal mutual impact on their radiation characteristics.} of dipole elements with half-wavelength spacing, leading to identical \ac{LOS} \ac{AoD} $\vartheta_{\rm LOS}$ and \ac{AoA} $\varphi_{\rm LOS}$ across the two bands.
We assume perfect synchronization in both time and frequency,  among all transmit and receive antennas. 
Furthermore, each antenna is connected to a dedicated \ac{RF} chain, allowing fully digital beamforming in both bands.
The considered multi-band system operates in \ac{TDD} mode, ensuring reciprocal channel responses.

\subsection{Multi-Band Channel Model} \label{subsec:multi_band_channel_model}
In this work, we consider channel estimation in the \ac{MMW} band by leveraging out-of-band information obtained from the sub-6\,GHz band.
Therefore, it is essential to follow an appropriate simulation methodology capable of generating frequency-dependent channels across multiple bands. 
In this work, we adopt the \ac{3GPP} channel model~\cite{3gpp.38.901} to ensure consistent and realistic multi-frequency simulations.
More specifically, the considered \ac{3GPP} channel model enables the joint generation of channels at different carrier frequencies while preserving the inherent spatial correlation of propagation paths. 

For a given delay tap $\tau$, the channel is represented by a complex-valued matrix $\NoLineBM{H}{b} [\tau] \in \mathbb{C}^{\TRx{M}{Rx}{b} \times \TRx{M}{Tx}{b}}$, based on the equivalent baseband formulation of an \ac{OFDM} system.
We consider a frequency-selective, time-invariant channel in the delay domain, modeled according to a Rician fading process as~\cite{molisch2012wireless}
\begin{equation}
    \begin{split}        
        \NoLineBM{H}{b} [\tau] & =
          \sqrt{\TRx{\eta}{}{b}} 
         \sqrt{\frac{\TRx{\kappa}{}{b}}{1+\TRx{\kappa}{}{b}}} 
         \underbrace{\NoLineBM{H}{b}_{\rm fs} \delta[0] }_{\rm free-space} \\
        & +
        \sqrt{\TRx{\eta}{}{b}}
        \sqrt{\frac{1}{ 1+\TRx{\kappa}{}{b}} } 
        \underbrace{\NoLineBM{H}{b}_{\rm sp} [\tau]}_{\rm stochastic\, part}.
    \end{split}       
    \label{eq:channel_model}
\end{equation}
In~\cref{eq:channel_model}, the parameter $\TRx{\eta}{}{b}$ represents large-scale effects such as path loss and shadow fading, as described in detail in~\cite{Pasic2025_TCOM}. 
The parameter $\TRx{\kappa}{}{b}$ denotes the Rician $K$-factor for the corresponding frequency band.
To capture the distinct propagation behavior across frequencies, the \ac{MMW} $K$-factor is modeled as a scaled version of its sub-6\,GHz counterpart, i.e., $\TRx{\kappa}{}{m}=c_{\kappa} \TRx{\kappa}{}{s}$, where $c_{\kappa}$ denotes the scaling factor.
Finally, the channel representation in the frequency domain for subcarrier $n$, denoted by $\NoLine{H}{b} [n]$, is obtained by applying the \ac{DFT} to the delay-domain channel $\NoLineBM{H}{b} [\tau]$.

The deterministic \ac{LOS} component is modeled according to a free-space far-field propagation assumption and is represented by
\begin{equation}
    \NoLineBM{H}{b}_{\rm fs} =
    e^{j \TRx{\chi}{}{b}}
    \TRx{\mathbf{a}}{Rx}{b} \NormalBrackets{ \varphi_{\rm LOS} }
    \NormalBrackets{\TRx{\mathbf{a}}{Tx}{b} \NormalBrackets{ \vartheta_{\rm LOS} }}^{\rm H},
    \label{eq:free_space_component}
\end{equation}
where
\begin{equation}
    \TRx{\mathbf{a}}{Tx}{b} \NormalBrackets{ \vartheta_{\rm LOS} } =  
    \begin{bmatrix}
        1 & \cdots  & 
        e^{-j 2 \pi \left( \TRx{M}{Tx}{b} - 1 \right) 
        \frac{ \TRx{\Delta d}{}{b}}{ \TRx{\lambda}{}{b}} 
        \sin \NormalBrackets{ \vartheta_{\rm LOS} } } 
        \end{bmatrix}^{\rm T}
        \label{eq:AoD_steering_vector}
\end{equation}
and
\begin{equation}
    \TRx{\mathbf{a}}{Rx}{b} \NormalBrackets{ \varphi_{\rm LOS} } =  
    \begin{bmatrix}
        1 & \cdots  & 
        e^{-j 2 \pi \NormalBrackets{ \TRx{M}{Rx}{b} - 1 } 
        \frac{\TRx{\Delta d}{}{b}}{\TRx{\lambda}{}{b}} 
        \sin \NormalBrackets{ \varphi_{\rm LOS} } }     
        \end{bmatrix}^{\rm T}
\end{equation}
denote the transmit and receiver array steering vectors, which correspond to the \ac{AoD} $\vartheta_{\rm LOS}$ and \ac{AoA} $\varphi_{\rm LOS}$, respectively.
Additionally, a random phase term $\TRx{\chi}{}{b} ~\sim~\mathcal{U}~\left( - \pi, \pi \right)$ is included in~\cref{eq:free_space_component} to account for the random phase offset at the first (reference) antenna element.

The stochastic part of the channel is given by the channel matrix $\NoLineBM{H}{b}_{\rm sp} [\tau]$, which is obtained by sampling its continuous-time counterpart, defined as
\begin{equation}
    \begin{split}
        \NoLineBM{H}{b}_{\rm sp} (\tau) = &
        \sum_{c=1}^{C} 
        \sum_{r_c=1}^{R_c^{\left( \rm b \right)}}
        \alpha_{r_c}^{\left( \rm b \right)}
        \delta \NormalBrackets{\tau - \tau_c - \tau_{r_c}^{\left( \rm b \right)}} \\
        & \times \TRx{\mathbf{a}}{Rx}{b} \NormalBrackets{ \varphi_c + \varphi_{r_c}^{\left( \rm b \right)} }
        \NormalBrackets{ \TRx{\mathbf{a}}{Tx}{b} \NormalBrackets{ \vartheta_c + \vartheta_{r_c}^{\left( \rm b \right)} } }^{\rm H}.         
    \end{split}
    \label{eq:stochastic_part}
\end{equation}
In~\cref{eq:stochastic_part}, $C$ represents the number of clusters. 
Each cluster is characterized by a mean delay $\tau_c$, a mean \ac{AoA} $\varphi_c$ and a mean \ac{AoD} $\vartheta_c$. 
Within each cluster, multiple propagation paths (or rays) are present, with $R_c^{\left( \rm b \right)}$ rays connecting the transmitter and receiver.
Each individual ray $r_c$ is associated with a complex path gain $\alpha_{r_c}^{\left( \rm b \right)}$, along with small perturbations around the cluster means, including a relative delay shift $\tau_{r_c}^{\left( \rm b \right)}$, a relative \ac{AoA} shift $\varphi_{r_c}^{\left( \rm b \right)}$ and a relative \ac{AoD} shift $\vartheta_{r_c}^{\left( \rm b \right)}$.
The generation of these parameters in~\cref{eq:stochastic_part} follows the \ac{3GPP} channel modeling framework, which adapts them according to the selected propagation scenario.

Furthermore, for multi-frequency channel generation described in~\cite{3gpp.38.901}, the \ac{3GPP} channel model ensures partial consistency across frequency bands. 
In particular, large-scale cluster characteristics (i.e., number of clusters $C$, mean delays $\tau_c$, mean angular parameters $\varphi_c$ and $\vartheta_c$) are shared between the sub-6\,GHz and \ac{MMW} frequency bands.
In contrast, small-scale properties, including cluster powers described as complex path gains $\alpha_{r_c}^{\left( \rm b \right)}$, as well as the intra-cluster variations (i.e., relative delay shifts $\tau_{r_c}^{\left( \rm b \right)}$, relative \ac{AoA} shifts $\varphi_{r_c}^{\left( \rm b \right)}$ and relative \ac{AoD} shifts $\vartheta_{r_c}^{\left( \rm b \right)}$) are frequency-dependent and are therefore generated independently for each frequency band.
Further details on the multi-band channel generation procedure can be found in the \ac{3GPP} channel model specification described in~\cite{3gpp.38.901}.

\subsection{Link Establishment} \label{subsec:link_establishment}
Establishing a reliable communication link involves two main phases: a training phase followed by a data transmission phase. 
During the training phase, the wireless channel is estimated at both sub-6\,GHz and \ac{MMW} frequency bands. 
These estimates are then utilized jointly to enable efficient data transmission over the \ac{MMW} link.

\subsubsection{Training Phase} \label{subsubsec:training_phase}
In the training phase, pilot symbols are transmitted isotropically in space over both frequency bands to acquire initial channel estimates. 
Each transmit antenna is assigned a set of pilot symbols that are known at the receiver.
The pilot symbols $\TRx{\boldsymbol{\phi}}{}{b} [n] \in \mathbb{C}^{\TRx{M}{Tx}{b} \times 1}$, drawn from a \ac{QAM} constellation, are distributed across $\TRx{N}{}{b}$ subcarriers such that different antennas occupy mutually exclusive (non-overlapping) subsets of subcarriers.
For the $t$-th transmit antenna, where $t \in \CurlyBrackets{ 1, \ldots, \TRx{M}{Tx}{b} }$, the pilot allocation is given by
\begin{equation}
    \mathrm{\phi}_t^{\left( \rm b \right)} [n] = 
    \begin{cases}
        \TRx{\mathrm{\phi}}{}{b} [n], \, n \in \{ t, t+\TRx{M}{Tx}{b}, \ldots, \TRx{N}{}{b} - \TRx{M}{Tx}{b} + t \}
        \\
        0, \quad \quad \, \, \, \text{else}
    \end{cases}
    \label{eq:pilot_allocation}
\end{equation}
The input-output relationship of the training phase is given by
\begin{equation} 
    \NoLine{y}{b} [n] = 
    \NoLine{H}{b} [n] 
    \TRx{\boldsymbol{\phi}}{}{b} [n] +
    \NoLine{w}{b} [n],
    \label{eq:training_input_output}
\end{equation}
where $\NoLine{y}{b} [n] \in \mathbb{C}^{\TRx{M}{Rx}{b} \times 1}$ represents the received signal and $\NoLine{w}{b} [n] \sim \mathcal{CN}\NormalBrackets{0,\sigma_{\TRx{w}{}{b}}^2 \mathbf{I}_{\TRx{M}{Rx}{b}}}$ denotes the \ac{AWGN} with the power of $\sigma_{\TRx{w}{}{b}}^2$.    
Based on the received pilot symbols, \ac{LS} estimates of the channel are first obtained at the pilot subcarrier locations. 
To reconstruct the channel over the entire bandwidth, linear interpolation is applied across the remaining subcarriers, yielding an estimated channel matrix $\Estimate{H}{b} [n] \in \mathbb{C}^{\TRx{M}{Rx}{b} \times \TRx{M}{Tx}{b}}$.
As the channel estimate is initially obtained at the receiver, it must also be made available at the transmitter to enable optimal beamforming design.
Since the system operates using the \ac{TDD} protocol, channel acquisition at the transmitter is achieved by exploiting channel reciprocity in both frequency bands.

\subsubsection{Data Transmission} \label{subsubsec:data_transmission}
In the data transmission phase, communication is carried out exclusively over the \ac{MMW} band.
The channel estimates obtained during the training phase are further processed using the methods, which will be introduced in~\cref{sec:methods}.
The resulting \ac{MMW} channel estimate $\BarX{H}{m} [n]$ is then used to design the precoder $\BarX{F}{m} [n]$ and combiner $\BarX{Q}{m} [n]$.
To maximize the \ac{SE} of the \ac{MIMO} system, we employ \ac{SVD} of the channel matrix.
The compact-form \ac{SVD} of $\BarX{H}{m} [n]$ is given by
\begin{equation}
    \BarX{H}{m} [n] = 
    \BarX{Q}{m} [n] \,
    \BarX{\Sigma}{m} [n]
    \NormalBrackets{\BarX{F}{m} [n]}^{\rm H},
    \label{eq:svd}
\end{equation}
where $\BarX{Q}{m} [n] \in \mathbb{C}^{\TRx{M}{Rx}{m} \times {\ell_{\rm max}}}$ and $\BarX{F}{m} [n] \in \mathbb{C}^{\TRx{M}{Tx}{m} \times {\ell_{\rm max}}}$ are semi-unitary matrices containing the left and right singular vectors, respectively.
The diagonal matrix $\BarX{\Sigma}{m} [n]$ consists of singular values, with each diagonal entry $\overline{\sigma}_{(i)}^{\left( \rm m \right)} [n]$ representing the $i$-th largest singular value of the channel.
Assuming a full-rank channel, the maximum number of spatial data streams is $\ell_{\rm max} = {\rm min} \NormalBrackets{ \TRx{M}{Rx}{m}, \TRx{M}{Tx}{m} }$. 
Power allocation across the parallel data streams is controlled by a diagonal matrix $\BarX{P}{m} [n]$.
To achieve the maximum achievable rate, the transmit power is distributed according to the water-filling principle~\cite{molisch2012wireless}. 
The total transmit power constraint
\begin{equation}
    \norm{ 
    \BarX{F}{m} [n]
    \NormalBrackets{\BarX{P}{m} [n]}^{1/2}
    }_F^2 = 
    \TRx{P}{\rm T}{m}
    \label{eq:power_normalization}
\end{equation}
is satisfied by the precoder for all subcarriers.

Let $\NoLine{x}{m} [n] \in \mathbb{C}^{\ell_{\rm max} \times 1}$ denote the transmitted symbol vector. 
The input-output relationship for the data transmission phase can be expressed as
\begin{equation} 
    \begin{split}
    \NoLine{y}{m} [n] & =
    \NormalBrackets{ \BarX{Q}{m} [n] }^{\rm H}
    \NoLine{H}{m} [n]
    \BarX{F}{m} [n]
    \NormalBrackets{ \BarX{P}{m} [n]}^{1/2}
    \NoLine{x}{m} [n] \\
    & + 
    \NormalBrackets{ \BarX{Q}{m} [n] }^{\rm H}
    \NoLine{w}{m} [n],
    \end{split}
    \label{eq:data_input_output}
\end{equation}
where $\NoLine{y}{m} [n] \in \mathbb{C}^{\ell_{\rm max} \times 1} $ is the received symbol vector. 
Owing to the semi-unitary property of $\BarX{Q}{m} [n]$, the transformed noise $ \NormalBrackets{ \BarX{Q}{m} [n] }^{\rm H} \NoLine{w}{m} [n]$ follows the same statistical distribution as $\NoLine{w}{m} [n]$.

\section{Channel Estimation Methods} \label{sec:methods}
In this section, we first revisit conventional in-band channel estimation methods for the \ac{MMW} system.
In particular, we describe the \ac{LS} estimator introduced in~\cref{subsubsec:training_phase}, followed by the \ac{LMMSE} estimator.
Subsequently, we then introduce the proposed out-of-band aided beam-domain channel estimation method.

\subsection{Least-Squares (LS)} \label{subsec:ls}
The \ac{LS} estimator serves as a low-complexity baseline for \ac{MMW} channel estimation. 
It relies solely on in-band observations and does not exploit any prior statistical information about the channel.
Specifically, an initial \ac{MMW} estimate $\Estimate{H}{m} [n]$ is obtained at pilot subcarriers using \ac{LS} estimation, followed by linear interpolation across frequency to obtain the channel on all subcarriers, as detailed in~\cref{subsubsec:training_phase}.
The resulting channel estimate $\BarX{H}{m} [n]$ is given by
\begin{equation}
    \BarX{H}{m} [n] = \Estimate{H}{m} [n]. 
    \label{eq:LS}
\end{equation}

\subsection{Linear Minimum Mean Squared Error (LMMSE)} \label{subsec:lmmse}
To improve estimation accuracy, we consider the \ac{LMMSE} approach which incorporates second-order channel statistics. 
This method uses the \ac{LS} estimate $\Estimate{H}{m} [n]$ as an initial input and refines it by incorporating the spatial covariance matrix of the channel and the noise variance.
The resulting estimate is given by
\begin{equation}
    \BarX{H}{m} [n] = 
    \Estimate{R}{m}_{\rm H}
    \NormalBrackets{ \Estimate{R}{m}_{\rm H} + \sigma_{\TRx{w}{}{m}}^2 \mathbf{I}_{\TRx{M}{Rx}{m}} }^{-1}
    \Estimate{H}{m} [n].
    \label{eq:LMMSE}
\end{equation}
In~\cref{eq:LMMSE}, $\Estimate{R}{m}_{\rm H}$ represents the estimated spatial covariance matrix, which is given by
\begin{equation}
    \Estimate{R}{m}_{\rm H} = 
    \frac{1}{\TRx{N}{}{m} \TRx{M}{Tx}{m}}
    \sum_{n=1}^{\TRx{N}{}{m}}
    \Estimate{H}{m} [n] 
    \NormalBrackets{ \Estimate{H}{m} [n] }^{\rm H}.
    \label{eq:spatial_covariance}
\end{equation}
In contrast to the \ac{LS} estimator, the \ac{LMMSE} estimator requires additional statistical information, namely the channel covariance and noise variance, and incurs higher computational complexity. 

\subsection{\underline{O}ut-of-\underline{B}and \underline{A}ided \underline{B}eam-Domain \underline{E}stimation (OBABE)}
This method leverages channel estimates from both the \ac{MMW} and sub-6\,GHz bands to improve channel estimation performance at \ac{MMW}.
In particular, the sub-6\,GHz out-of-band channel estimate is exploited to identify the dominant angular components of the propagation channel, which are then employed as prior information to filter out low-power and noise-dominated components in the \ac{MMW} beam-domain estimate.
The overall procedure is described as follows.

We first transform the frequency-domain channel estimates $\Estimate{H}{b}$, for both-frequency bands $b$, into the beam domain.
Assuming \acp{ULA}, the beam-domain representation is given by
\begin{equation}
    \EstimateBeam{H}{b} [n] =
    \frac{1}{\sqrt{\TRx{M}{Tx}{b} \TRx{M}{Rx}{b}} }
    \left( \TRx{\mathbf{U}}{Rx}{b} \right) ^ {\rm H} 
    \Estimate{H}{b} [n]
    \TRx{\mathbf{U}}{Tx}{b}, 
    \label{eq:freq_to_beam_domain}
\end{equation}
where the matrices $\TRx{\mathbf{U}}{Tx}{b} \in \mathbb{C}^{ \TRx{M}{Tx}{b} \times \TRx{M}{Tx}{b} } $ and $\TRx{\mathbf{U}}{Rx}{b} \in \mathbb{C}^{ \TRx{M}{Rx}{b} \times \TRx{M}{Rx}{b} }$ are \ac{DFT} matrices, respectively.
These \ac{DFT} matrices consist of orthogonal steering vectors that span the angular domain, enabling a spatial decomposition of the channel into distinct beam directions.

Next, we average the sub-6\,GHz beam-domain estimates across subcarriers, which is given by
\begin{equation}
    \HatBeam{H}{s} = 
    \frac{1}{\TRx{N}{}{s}}
    \sum_{n=1}^{\TRx{N}{}{s}}
    \EstimateBeam{H}{s} [n],
    \label{eq:beam_averaging_s}
\end{equation}
where $\HatBeam{H}{s} \in \mathbb{C}^{\TRx{M}{Rx}{s}\times\TRx{M}{Tx}{s}}$ denotes the averaged beam-domain estimate.
This averaging reduces frequency-dependent fluctuations and yields a stable representation of the dominant spatial structure.
However, since the antenna dimensions at \ac{MMW} are typically equal to or larger than those at sub-6\,GHz, the resulting reference must be adapted accordingly.
To this end, the averaged sub-6\,GHz beam-domain estimate is upsampled to match the \ac{MMW} antenna configuration, resulting in
\begin{equation}
    \HatBeam{H}{m} = \HatBeam{H}{s} \otimes \mathbf{1}_{c_{\rm Rx}\times c_{\rm Tx}}.
\end{equation}
The upsampling factors $c_{\rm Tx} = \TRx{M}{Tx}{m} / \TRx{M}{Tx}{s}$ and $c_{\rm Rx} = \TRx{M}{Rx}{m} / \TRx{M}{Rx}{s}$ account for the difference in the number of transmit and receive antennas between the \ac{MMW} and sub-6\,GHz systems, respectively.
This ensures that both representations are defined over the same beam-space grid, enabling direct comparison and filtering.
In practical deployments, antenna array sizes are typically chosen as powers of two. 
Accordingly, we assume that the sub-6\,GHz antenna configuration is an integer divisor of the corresponding \ac{MMW} configuration.
Next, we vectorize the upsampled reference beam-domain matrix, given by
\begin{equation}
    \HatBeam{h}{m} = 
    \vectorize{ \HatBeam{H}{m} } \in \mathbb{C}^{\TRx{M}{}{m} \times 1},
    \label{eq:beam_vectorize_s}    
\end{equation}
where $\TRx{M}{}{m} = \TRx{M}{Tx}{m} \TRx{M}{Rx}{m}$ represents the total number of \ac{MMW} beam-domain coefficients. 
Further, we calculate the power contribution of each beam and normalize it as $\HatBeam{p}{m} = \abs{ \HatBeam{h}{m} }^2 / \norm{ \HatBeam{h}{m} }^2$.
To identify dominant beam components, we sort $\HatBeam{p}{m}$ in descending order and then compute the corresponding cumulative power distribution, yielding
\begin{equation}
    \HatBeam{c}{m} =
    \NoLine{L}{m}
    \NoLine{\boldsymbol{\Pi}}{m}
    \HatBeam{p}{m},
    \label{eq:beam_cumsum}    
\end{equation}
where $\NoLine{\boldsymbol{\Pi}}{m} \in \mathbb{R}^{ \TRx{M}{}{m} \times \TRx{M}{}{m}}$ represents the permutation matrix which sorts the elements of $\HatBeam{p}{m}$ in descending order and
\begin{equation}
    \NoLine{L}{m} = 
    \begin{bmatrix}
    1 & 0 & \cdots & 0 \\
    1 & 1 & \cdots & 0 \\
    \vdots& \vdots & \ddots & \vdots \\
    1 & 1 & \cdots & 1
    \end{bmatrix}
    \in \mathbb{R}^{ \TRx{M}{}{m} \times \TRx{M}{}{m}}
    \label{eq:lower_triangular}    
\end{equation}
denotes the lower triangular matrix that performs the cumulative summation.
Based on this, we determine the minimum number of dominant beams $\widehat{i}_B$, given by
\begin{equation}
    \widehat{i}_B = 
    \min \CurlyBrackets{ i \in \CurlyBrackets{1, \ldots, \TRx{M}{}{m}} : \widehat{c}_{B, i}^{\left( \rm m \right)} > p},
\end{equation}
whose cumulative power accounts for a fraction $p \in \SquareBrackets{0, 1}$ of the total power.
A higher value of $p$ retains more \ac{MMW} beams, while a lower value results in stronger beam sparsification.
We then extract the corresponding index set $\widehat{\mathcal{I}}_B$ of the $\widehat{i}_B$ dominant beams from $\HatBeam{p}{m}$, given by
\begin{equation}
    \widehat{\mathcal{I}}_B = 
    \operatorname*{arg\,top}_{\displaystyle \widehat{i}_B}\,    \NormalBrackets{ \HatBeam{p}{m} }.
\end{equation}
This index set is used to filter the \ac{MMW} beam-domain channel estimate.
Specifically, we vectorize the instantaneous \ac{MMW} beam-domain channel for each subcarrier by
\begin{equation}
    \EstimateBeam{h}{m} [n] = 
    \vectorize{ \EstimateBeam{H}{m} [n] }
\end{equation}
and retain only entries indexed by $\widehat{\mathcal{I}}_B$, while setting all other components to zero, as follows:
\begin{equation}
    \overline{\mathrm{h}}_{B, i}^{\left( \rm m \right)} [n] =
    \begin{cases}
        \widetilde{\mathrm{h}}_{B, i}^{\left( \rm m \right)} [n], \, \quad i \in \widehat{\mathcal{I}}_B
        \\
        0, \quad \quad \, \, \, \quad \text{else}
    \end{cases}
\end{equation}
Finally, we reshape the vectorized estimate back into the matrix form by
\begin{equation}
    \BarBeamX{H}{m} [n] = 
    \unvectorize{ \BarBeamX{h}{m} [n] }
\end{equation}
and subsequently transform the beam-domain channel estimate back to the frequency domain, yielding the resulting channel estimate
\begin{equation}
    \BarX{H}{m} [n] =
    \frac{1}{\sqrt{\TRx{M}{Tx}{m} \TRx{M}{Rx}{m}} }
    \TRx{\mathbf{U}}{Rx}{m} \, 
    \BarBeamX{H}{m} [n] \, 
    \NormalBrackets{\TRx{\mathbf{U}}{Tx}{m}}^{\rm H}. 
    \label{eq:beam_to_freqdomain}
\end{equation}

\section{Simulation-based Comparison} \label{sec:simulations}
To assess the performance of the proposed channel estimation methods, we simulate the achievable \ac{SE} over a frequency-selective channel.
The key simulation parameters are listed in~\cref{tab:simParams}.
For modeling the frequency-selective channel, we adopt the \ac{3GPP} model for the indoor-office scenario with an adjustable $K$-factor, following the methodology described in~\cite{3gpp.38.901}.
Due to the strong similarity between $K$-factors across bands~\cite{Pasic2025}, we set the scaling parameter to $c_\kappa=1$ (i.e., 0\,dB), such that the \ac{MMW} $K$-factor satisfies $\TRx{\kappa}{}{m}= \TRx{\kappa}{}{s}$.
As a reference, we first analyze conventional \ac{LS} and \ac{LMMSE} channel estimation methods that rely solely on \ac{MMW} observations, as introduced in~\cref{subsec:ls} and~\cref{subsec:lmmse}.
In addition, we include the performance with perfect \ac{CSI}, i.e., $\BarX{H}{m} [n] = \NoLine{H}{m} [n]$.

\begin{table}[t]
    \centering
    \caption{Simulation Parameters} 
    \label{tab:simParams}
    \begin{tabular}{rcc}
        \hline
        \textbf{Parameter}                          & \multicolumn{2}{c}{\textbf{Value}} \\ \hline
        Frequency Band                              & sub-6 GHz         & mmWave         \\
        Carrier Frequency $f_{\rm c}$               & 2.55\,GHz         & 25.5\,GHz           \\
        Wavelength $\lambda$                        & 11.76\,cm         & 1.176\,cm          \\
        Bandwidth $B$                               & 20.16\,MHz         & 403.2\,MHz            \\
        Subcarrier Spacing $\bigtriangleup f$       & 60\,kHz           & 240\,kHz             \\
        Antenna Configuration $\NormalBrackets{ M_{\rm Rx} \times M_{\rm Tx}}$    & 8$\times$8               & 16$\times$16 \\           
        Transmit Power $P_{\rm T}$                  & 30\,dBm           & 30\,dBm \\ 
        Noise Figure $F$                            & 3\,dB             & 6\,dB            \\    
        Fraction $p$                                & N/A               & 0.95      \\
        Number of Realizations $L_r$                & 5000              & 5000             \\  \hline
    \end{tabular}
\end{table}

We consider the achievable \ac{SE} expressed in bits$/$s$/$Hz as the main performance metric.
Since this work focuses on the impact of channel estimation quality on achievable data rates and all considered methods employ identical training overhead, the training overhead is not included in the \ac{SE} calculation.
Throughout the analysis, we assume equal power allocation across all subcarriers.
The achievable \ac{SE} averaged over $\TRx{N}{}{m}$ subcarriers is given by
\begin{equation} 
    \mathrm{SE}= 
    \frac{1}{\TRx{N}{}{m}} 
    \sum_{n=1}^{\TRx{N}{}{m}} 
    \sum\limits_{\mu=1}^{\ell_{\rm max}} 
    \log_2 \NormalBrackets{ 1 + \mathrm{SINR}_{\rm \mu} [n] },
    \label{eq:se}
\end{equation}
where the effective \ac{SINR} for the stream ${\rm \mu}$ is denoted by~\cite{Kammoun2014}
\begin{equation} 
    \mathrm{SINR}_{\rm \mu} [n] = \frac{ 
    \abs{ \TRx{ \overline{\mathrm{G}}}{\rm \mu,\mu}{m} [n]}^2 }
    { \sum\limits_{\substack{\nu=1 \\ \nu \neq \mu}}^{\ell_{\rm max}} 
    \abs{ \TRx{\overline{\mathrm{G}}}{\rm \mu, \nu}{m} [n] }^2
    + \NormalBrackets{ \sigma_{\rm \mu}^2 + \sigma_{\TRx{w}{}{m}}^2 } 
    \norm{ \TRx{\overline{\mathrm{Q}}}{\rm :,\mu}{m} [n]}^2}.    
    \label{eq:sinr}
\end{equation}
The elements $\TRx{ \overline{\mathrm{G}}}{\rm \mu,\nu}{m} [n]$, with $\mu, \nu \in \{ 1, \ldots, \ell_{\rm max} \}$, denote the entries of the channel gain matrix $\BarX{G}{m} [n] \in \mathbb{C}^{\ell_{\rm max} \times \ell_{\rm max}}$ for the $n$-th subcarrier, which is given by
\begin{equation}
    \BarX{G}{m} [n]= 
    \NormalBrackets{ \BarX{Q}{m} [n] }^{\rm H}
    \NoLine{H}{m} [n] \,
    \BarX{F}{m} [n]
    \NormalBrackets{ \BarX{P}{m} [n]}^{1/2}.
    \label{eq:chgain}
\end{equation}
Moreover, $ \sigma_{\rm \mu}^2$ corresponds to the diagonal elements (variance) of the estimation error covariance matrix $\BarX{C}{m}_{\varepsilon} \in \mathbb{C}^{\ell_{\rm max} \times \ell_{\rm max}}$ averaged over subcarrier, given by
\begin{equation}
    \BarX{C}{m}_{\varepsilon} =
    \frac{1}{\TRx{N}{}{m} \ell_{\rm max}}
    \sum_{n=1}^{\TRx{N}{}{m}} 
    \TRx{\overline{\boldsymbol{\varepsilon}}}{}{m} [n]
    \NormalBrackets{ \TRx{\overline{\boldsymbol{\varepsilon}}}{}{m} [n] }^{\rm H}.
    \label{eq:error_covariance}
\end{equation}
The estimation error matrix is given by
$    \TRx{\overline{\boldsymbol{\varepsilon}}}{}{m} [n] =
    \NoLine{G}{m} [n] - 
    \BarX{G}{m} [n]. $
Finally, $\TRx{\overline{\boldsymbol{\varepsilon}}}{}{m} [n]$ represents the mismatch between the channel gain matrix obtained using the estimated precoder/combiner and the one obtained under perfect \ac{CSI}.

\begin{figure}[t]
    \centering
    {\includegraphics[width=0.96\columnwidth]{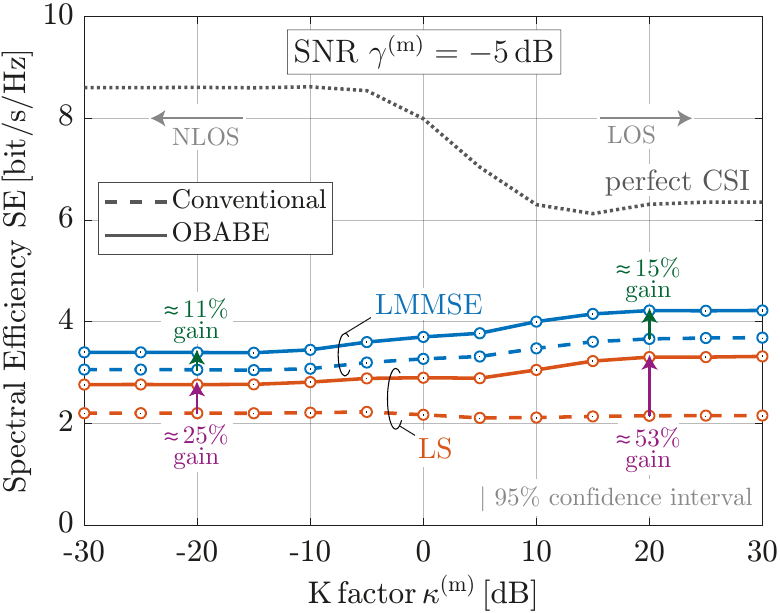}}
    \caption{
    Incorporating out-of-band information via the proposed OBABE method brings consistent improvements over conventional in-band estimation, irrespective of the $K$-factor or the baseline estimator.
    }
    \label{fig:SE_vs_K_factor}
\end{figure}

We evaluate the performance of the proposed channel estimation methods as a function of the Rician $K$-factor.
The simulation results are presented in terms of achievable \ac{SE} in~\cref{fig:SE_vs_K_factor}.

As expected, channel estimation based on \ac{LS} generally yields inferior \ac{SE} performance compared to \ac{LMMSE}-based estimation due to the absence of statistical prior information.
However, when incorporating out-of-band information via the proposed \ac{OBABE} method, a consistent \ac{SE} improvement is observed over conventional in-band estimation, irrespective of the $K$-factor or the baseline estimator.
For the \ac{LS}-based implementation, \ac{OBABE} provides substantial gains over its in-band counterpart. 
In particular, the \ac{SE} improves by approximately 25\% in the low $K$-factor regime ($-$20\,dB, \ac{NLOS} conditions) and up to 53\% in the high $K$-factor regime (20\,dB, \ac{LOS} conditions).
In contrast, when \ac{OBABE} is applied on top of \ac{LMMSE} estimation, the relative gains are more moderate, as the \ac{LMMSE} baseline already achieves strong performance. Specifically, the \ac{SE} gain is around 11\% for a low $K$-factor of $-$20\,dB and increases to approximately 15\% for a high $K$-factor of 20\,dB.

Furthermore, we observe higher gains in the \ac{LOS} regime (i.e., at higher $K$-factor). 
The primary reason for this behavior is the increased spatial sparsity under \ac{LOS} conditions, where the channel energy is predominantly concentrated in a small number of beam components, typically dominated by the direct path.
In this case, the number of selected dominant beams is approximately $\widehat{i}_B=35$, corresponding to only 14\% of the total beams.
In contrast, in the \ac{NLOS} regime (lower $K$-factors), the channel exhibits reduced sparsity, with energy distributed across multiple channel paths. 
Consequently, the number of selected dominant beams increases to approximately $\widehat{i}_B=113$, corresponding to 44\% of the total beams.
Moreover, the dominant propagation paths are less consistent across frequencies, which limits the predictive capability of sub-6\,GHz information for \ac{MMW}.

\section{Conclusion} \label{sec:conclusion}
In this paper, we propose a novel method for beam-domain \ac{CSI} estimation in \ac{MMW} \ac{MIMO} systems by exploiting out-of-band beam-domain \ac{CSI} obtained from the sub-6\,GHz band.
Numerical results demonstrate the effectiveness of the proposed  method compared to conventional approaches that rely solely on in-band \ac{MMW} information, achieving consistent performance gains in both \ac{NLOS} and \ac{LOS} channel conditions.

\bibliography{references}
\bibliographystyle{IEEEtran}

\end{document}